\documentclass[prb,showpacs,showkeys,preprintnumbers,superscriptaddress,a4paper,twocolumn]{revtex4-1}
\usepackage[T1]{fontenc}
\usepackage[utf8]{inputenc}
\usepackage{graphicx}
\usepackage{amsmath,amssymb}
\usepackage{amstext}
\usepackage{lipsum}
\usepackage{mathtools}
\usepackage{afterpage}
\usepackage{amssymb}
\usepackage{epsfig}
\usepackage{epstopdf} 
\usepackage{filecontents}
\usepackage[colorlinks,citecolor=blue,filecolor=blue,linkcolor=blue,urlcolor=blue]{hyperref}

\newcommand{\ex}{\mathbf{e}_x}
\newcommand{\ez}{\mathbf{e}_z}
\newcommand{\etx}{\tilde{\mathbf{e}}_x}
\newcommand{\etz}{\tilde{\mathbf{e}}_z}
\newcommand{\ety}{\tilde{\mathbf{e}}_y}

\def\beq{\begin{equation}}
\def\eeq{\end{equation}}
\def\bea{\begin{eqnarray}}
\def\eea{\end{eqnarray}}
\def\ba{\begin{array}}
\def\ea{\end{array}}

\begin{document}
\title{Classical ground states, spin-wave and PCUT analysis of $\rm H_2SQ$ system}
\author{Vikas Vijigiri}\email{vikasvikki@iopb.res.in}\affiliation{Institute of Physics, Bhubaneswar-751005, Orissa, India}
\affiliation{Homi Bhabha National Institute, Mumbai - 400 094, Maharashtra, India}
\author{Saptarshi Mandal }\email{saptarshi@iopb.res.in}\affiliation{Institute of Physics, Bhubaneswar-751005, Orissa, India}
\affiliation{Homi Bhabha National Institute, Mumbai - 400 094, Maharashtra, India}


\begin{abstract}
We study an organic Hydrogen bonded material $\rm{H_2SQ}$ analytically and map out the phase diagram  as well as low energy excitations in the relevant
parameter space. At zeroth order the dynamics is governed by plaquette interaction (product of $\sigma_z$ over a plaquette) which defines
a $Z_2$ gauge theory and a deconfinement phase satisfying "ice rules". The system is studied under additional interactions such as an external
Zeeman field (with strength $K$) in $x$-direction, a inter-molecular interaction (with strength $J_1$) and a dipole-dipole interaction with 
strength $J_2$ such that $K>J_1>J_2$. The effect of dipole-dipole interaction removes the local $Z_2$ symmetry and gives rise to four global
degenerate states with Ferroeletric order. Using meanfield analysis we chart out the phase diagram for classical version of the model and find
$K_c$ which defines the transition from disordered phase to ordered phase in $J_1, K$ plane for various values of $J_2$. We find that presence
of $J_1$ and $J_2$ tends to stabilize the deconfined phases. Over the classical ground states we perform spin-wave analysis and surprisingly find
that quantum fluctuations does not remove the classical degeneracy at all at quadratic level. The spin-wave spectrum is found for the four global
degenerate ground states which shows both  gapped spectrum  and gapless spectrum (near the vicinity of CDT transition) for $J_2=0$. However for $J_2$ finite, the spectrum is always gapped. We perform PCUT analysis to improve the results of spin-wave analysis and calculate ground state energy and one particle dispersion and gap at high symmetry point. Using this we draw the phase boundary between confined and deconfined phase in the $K-J_1$ plane. The effect of $J_2$ is also discussed in the resulting phase boundary. 
\end{abstract}

\date{\today}

\maketitle
\section{Introduction}

The hydrogen-bonded systems continue to serve as a basis of realizing  quantum effects at macroscopic scale \cite{J. C. Slater,G. A. Samara,P. S. Peercy} , as these are the systems which exhibit quantum tunneling(of positively charged Hydrogen atoms) between two adjacent sites. The position of Hydrogen atoms are usually replaced by the pseudo-spin 1/2 variable which then can be mapped into suitable quantum spin systems, for example water spin ice system~\cite{Anderson-1956}. Among the various hydrogen bonded materials, the organic $\rm{H_2SQ}$ is notable for their ferroelectric properties, light weightiness and eco friendly nature.  The ferroelectric properties of such system was first studied by Slater on ${\text{KH}_{2}\text{PO}_4}$(KDP) \cite{J. C. Slater}. Later several experimental observations of the ferroelectric properties were observed in several hydrogen-bonded ferroelectrics \cite{G. A. Samaraa,G. A. Samara,P. S. Peercy,Y. Moritomo} . Recently there has been effort to provide a  theoretical framework of these systems in the language of  quantum spin systems\cite{N. Shannon_1, N. Shannon,Chyh-Hong Chern}.  To minimize free energy the position of Hydrogen atoms in these systems are inherently subjected to the constraint called "ice rules", which constraints exactly two hydrogen atoms are approaching oxygen atom out of four hydrogen atoms\cite{J. C. Slater,L. Pauling,R. Savit,H.-D. Maier}. Squaric acid molecule ${\text{H}_{2}\text{SQ}}$ is a hydrogen bonded organic system which has a quasi two dimensional antiferroelectric layer where the square molecule SQ (say type A) is surrounded by four molecules of type B with hydrogen bonds \cite{Y. Moritomo}. This constraint is very much similar to the well known two-in and two-out configurations in Spin ice pyrochlores and other frustrated systems~\cite{castelnovo-2012}. Such systems with the ice rule constraints are shown to exhibit no long range ordering (LRO) down to zero temperature  with exotic magnetic monopoles as the elementary excitations \cite{N. Shannon,K. A. Ross,O. F. Syljuasen,E. Ardonne,N. Shannon_1}. The ice rules which are seen as the frustration in the system where all the configurations in the low energy sector are degenerate was discussed by Pauling in the early 1930s\cite{L. Pauling}.  While the quantum effects of the water ice and spin ice remains to attract us, the recent studies on water ice systems have shown to exhibit interesting phenomenon of coherent quantum tunneling in the low temperatures leading to U(1) Quantum Spin-liquid ground states with fractionalized spinon and gauge field excitations\cite{N. Shannon,K. A. Ross,O. F. Syljuasen,E. Ardonne,N. Shannon_1,C. Castelnovo}. 

\vspace{0.18cm}

Here, in this paper we theoretically analyze  the phase diagram and low lying excitations of $\rm{H_2SQ}$ systems in general at zero temperature. The earlier work~\cite{Y. Moritomo} investigated the finite temperature phase diagram in the $T-P$ plane where $T$ is temperature and $P$ is pressure. They have found a phase transition from AFE (anti-ferroelectric) to PE (paraelectric ) phase. The paraelectric phase sustains at zero temperature above some critical pressure $P_c$. To our knowledge the first model Hamiltonian was proposed in Ref[\onlinecite{H.-D. Maier}] which view the $\rm{H_2SQ}$ is an interacting one-dimensional Ising spin chains.  Recently Chern et al\cite{Chyh-Hong Chern}  has studied similar  model Hamiltonian for such a aystem and obtained  phase diagram  numerically using quantum monte carlo technique at zero temperature. The Model Hamiltonian at zeroth order consists a four-spin plaquette interaction with a strength $J_0$. This interaction give rise to so called "ice-rules" and defines a deconfined phase. When the Hamiltonian is studied in the presence of a  magnetic field characterized by a parameter $K$, the system shows a confinement-deconfinement phase transition. The model also include a next-nearest neighbouring Ising like interaction with a strength $J_1$ and a dipole-dipole interaction $J_2 < J_1$. Usually the presence of $J_2$ causes the Ferroelectricity of the materials.  The model was shown to exhibit both confinement-deconfinement (CDT) and ferroelectric quantum phase transition (FT) for appropriate set of parameters of the model Hamiltonian.  However, in this work we carry out analytical calculations  which has not been done so far.  We have used three different complimentary  techniques, mainly classical meanfield analysis, spin-wave approximations over classical ground states and PCUT (perturbative continuous unitary transformation\cite{F. Wegnero,F. Wegner,G. S. Uhrig,C. Knetter,C. Knettero} ) to improve the results obtained in spin-wave approximations to obtain a complete theoretical understanding of the model.  Our plan of presentation is as follows. In section \ref{model}, we introduce the model Hamiltonian and explain various terms present. Next,  we analyze the classical version of the model Hamiltonian in section \ref{section-1} following which the spin wave excitations are obtained next. The PCUT analysis has been presented in section \ref{section-3}. The mapping of our model Hamiltonian to that of Kitaev's Toric code Hamiltonian~\cite{kitaev-2003} is discussed in section \ref{section-4} and we summarize our results at the end.

\section{Model}
\label{model}
We consider the following Hamiltonian given below\cite{Chyh-Hong Chern}:
\begin{equation}
H= H_0 + H_1 + H_2
\label{ham}
\end{equation}
where $H_0, H_1$ and $H_2$ are respectively  given by
\begin{eqnarray}
\label{hzero}
H_{0} & = & J_0 \sum_p \sigma_1^z\sigma_2^z\sigma_3^z\sigma_4^z - K \sum_i \sigma_i^x \\
\label{hone}
H_{1} & = &  J_1 \sum_p (\sigma_1^z\sigma_3^z + \sigma_2^z\sigma_4^z) \\
\label{htwo}
H_{2}& = &   - J_2 \sum_{\langle AB \rangle} \vec{P}_A\cdot \vec{P}_B    
\end{eqnarray}
where $\sigma^{\alpha,}_{i}$s are the Pauli matrices, $P_{A}$ and $P_{B}$ are the dipole moment vectors of molecule A and B respectively and the components of them are defined as follows,

\begin{eqnarray}
P_{(A,B)x} & = &(\pm) \frac{1}{4}(\sigma^{z}_{1} + \sigma^{z}_{2} - \sigma^{z}_{3} - \sigma^{z}_{4}) \\
P_{(A,B)y} & = & (\pm)\frac{1}{4}(\sigma^{z}_{2} + \sigma^{z}_{3} - \sigma^{z}_{1} - \sigma^{z}_{4})
\end{eqnarray} 

where (+) is for molecule A and (-) is for molecule B. The summation of indices 'p' runs over all the plaquettes of the dual lattice(red) and $i$ runs over all the spins in the dual lattice and $\langle AB \rangle$ indicates the nearest neighbor dipole-dipole interaction.
\begin{figure}[h]
\label{squareacid}
\includegraphics[width=9cm,height=15cm,keepaspectratio]{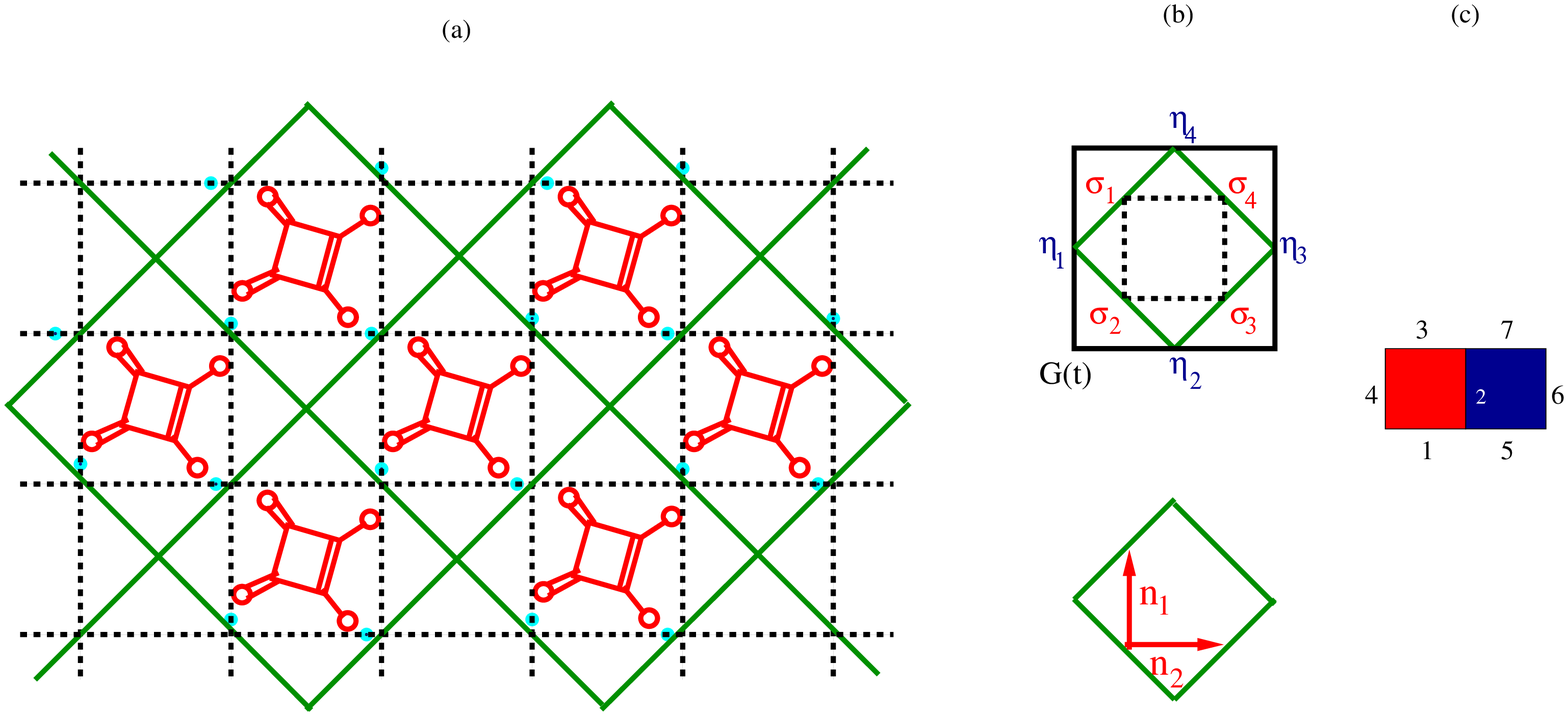}
\caption{The blue small dots represent the hydrogen atoms and the red circles represent the oxygen atoms of the molecule $\text{H}_{2}\text{SQ}$ forming a quasi-2D configuration of hydrogen-bonded network. The top right corner showing the $\eta$ variables defined on the dual lattice(green line) sites and the pseudo-spin variables $\sigma$ which are defined on the bonds of the lattice(black dashed line).} 
\end{figure}

The parameter space of the Hamiltonian $H$ is three dimensional with $J_0$ being the largest followed by $K, J_1$ and $J_2$ in magnitude. The ice rules were accounted in the form of $\mathcal{Z}_{2}$ gauge-invariant term $J_0(>0)$ and the intermolecular coupling term $J_1$ were generating a macroscopic degeneracy  accounted in the form of Ising ferromagnet. Since there are ground states with each square plaquette having finite dipole moments, a more general Hamiltonian would also have these dipole-dipole interaction term $J_2$. 


\section{Classical model}
\label{section-1}
In our endeavor to understand the different aspect of the complex Hamiltonian as given in Eqs. ~\eqref{ham} - \eqref{htwo}, we first examine the classical equivalent of the model where
the quantum spins are now replaced by the classical Heisenberg  spins.  For sake of conveniences we rewrite the various terms in Eqs. ~\ref{hzero}- \ref{htwo} as follows,
\begin{eqnarray}
H_{0}& = &-\frac{1}{4} J_0\sum_{\langle ii_{1}i_{2}i_{3}\rangle} S_{i_1}^zS_{i_2}^zS_{i_3}^zS_{i_4}^z - K \sum_i S_i^x \\
H_{1}& = & \frac{1}{2} J_{1} \sum_p (S_1^zS_3^z + S_2^zS_4^z) \\
H_{2}& = &  -\frac{1}{2} J_{2} \sum_{\langle AB \rangle} \vec{P}_A\cdot \vec{P}_B    
\end{eqnarray}
Following earlier study \cite{Chyh-Hong Chern}, we assume that $J_{0}$, $J_{1}$, $J_{2}$ > 0 and the prefactors $\frac{1}{4}$, $\frac{1}{2}$, $\frac{1}{2}$ account for the multiple counting of the the same term in the Hamiltonian.  \\

Let us briefly discuss the consequences of various terms in the above Hamiltonians.  The first terms in $H_0$ can  easily be satisfied by suitably aligning spins along $\pm$ z-axis such
that each plaquette exactly contain two up-spins and two down spins which are equivalent to the celebrated ice-rules. If one considers a torus having $\mathcal{N}_x$ and $\mathcal{N}_y$ plaquette  in $x$ and $y$ direction respectively then
there are $\mathcal{D}= 2^{\mathcal{N}_x \mathcal{N}_y}$ degenerate ground states for the first terms in the Hamiltonian. The 2nd largest scale problem i.e $K$ brings in frustration in the
tendency to align along the $\pm$ z-direction. As large $K$, all the spins are eventually aligned along the $x$-directions. One of our motivation to understand the classical version of the model was to investigate this transitions from highly degenerate classical configuration to a ordered phase where spins are along $x$-axis.   The terms in  $H_1$ and $H_2$ brings in additional complexity mainly by lifting the  ground state degeneracy generated by first term in $H$ partially.  The term $H_1$ reduces the degeneracy $ \mathcal{D}$ to $\tilde{\mathcal{D}}=2^{\mathcal{N}_x \mathcal{N}_y/2}$ and dipole-dipole interaction eventually removes all local degeneracy causing a Ferroelectric alignment of electric dipoles associated with each plaquette. However there are four global degenerate states in this case which will be discussed later in detail.\\

Generally for classical Heisenberg type interactions, one resorts to Luttinger-Tisza method ~\cite{luttinger-1946} to find the classical ground states. However the limitation of this approach is confined to only Bravais lattices, though for non-Bravais lattices, it may give important leads to possible ground state spin configurations ~\cite{mandal-2013}. The presence of four-spin interactions limits us from  using such analytical methods. Owing to this reason we examine numerically the ground state spin configurations.  We notice that the Hamiltonian can be re-written in the following form~\cite{sklan-2013} $H = \sum_i h^z_i S^z_i +  h^x_i S^x_i$  where for a given spin-component $S^{\alpha}_i$, $h^{\alpha}_i$ denotes the local field component along $\alpha$-axis. The minimum  energy configuration of spins are then obtained by aligning $S^{\alpha}_i$ to negative $\alpha$-axis. Usually one starts from a random configurations of $[S^{\alpha}_{i,0}]$ yielding a configurations of $[h^{\alpha}_{i,0}]$ and a total energy $E[S^{\alpha}_{i,0}]$. The distribution $[h^{\alpha}_{i,0}]$ yields a new configurations of spins  $[S^{\alpha}_{i,1}]$ and new total energy of the system $E[S^{\alpha}_{i,1}]$. In the above the index `$0$' and `$1$' denotes the steps in numerical iterations. We continue this process until $E[S^{\alpha}_{i,n}] \equiv E[S^{\alpha}_{i,n+1}]$. We have performed numerical simulations over lattice of dimension $256 \times 256$ and checked for sufficient initial configurations. Surprisingly we have found that the ground state has a one to one corresponds to the ground state configurations of the first terms of $H_0$. The only difference is that the spins has now a finite and constant value of $S^{x}_i$ which changes as a function of $K$ and other parameters similar to what has been found for Kitaev model in the presence of transverse magnetic field in an earlier study~\cite{vidal-2008}. Thus the ground state configurations can be written as,

\begin{eqnarray}
\label{spincon}
\vec{S}_i &=& S \left( \lambda_i  \cos \theta   \ez +  \sin \theta \ex \right)
\end{eqnarray}
where $\lambda_i$ could be $\pm$ in tune with the ground state configurations of $H_0$ for $K=0$.  The value of $\theta$ depends on $K, J_1, J_2$. For $K=0$, we have $\theta= 0$. This $\theta$ takes the role of our order paremeter. From the meanfield results represented by Eqn \ref{spincon}, the ground state energy of the system can be written as follows,
\begin{eqnarray}
\label{clasen}
E_{cl}& = & -\frac{1}{2}J_{0}S^4N\cos^{4}\theta - J_{1}S^2N\cos^{2}\theta \nonumber  \\
      & & -  KSN\sin\theta -2J_{2}S^2N\cos^{2}\theta
\end{eqnarray}

\begin{figure*}
\label{contour}
\includegraphics[height=45mm,width=180mm]{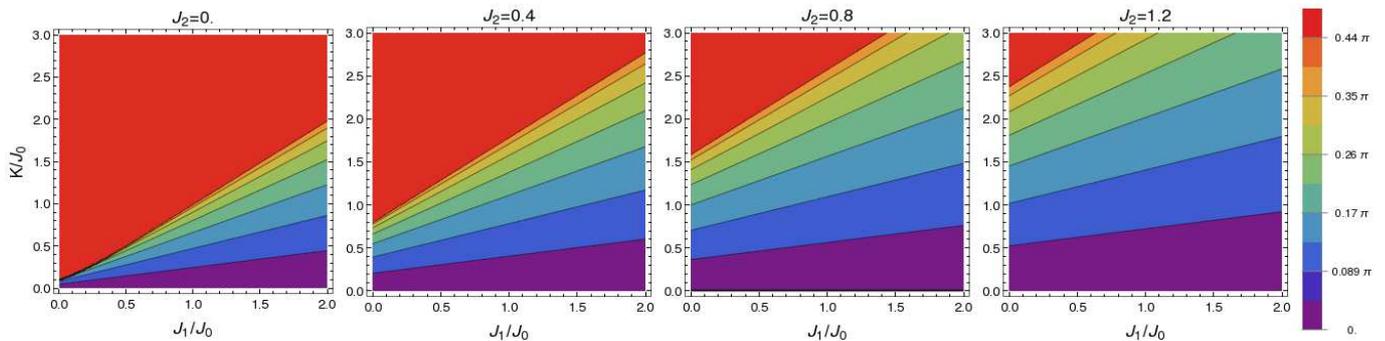}
\caption{In the above figure we have represented contour plot of $\theta_c$ in the $K-J_1$ plane for various values of $J_2$. The red region denotes ordered phase where all the spins align in $x$-direction. Various other shaded region (except red) denotes a disordered phase where the $z$-competent of
spins are disordered.} 
\end{figure*}

Minimizing $E_{cl}$ with respect to $\theta$ we obtain $\theta_{C}$ which minimizes the ground state energy $E_{cl}(\theta_{C})$. This ground state energy has been compared with the $E_{x}= - K N$ which denotes the energy corresponding to the state where all spins are aligned along $x$-direction.  For a given $J_0, J_1, J_2$ there exists  a  $K_c$ such that  if $K \leq K_c$ then $E_{cl}  < E_{x}(\theta_C)$ with $\theta_C \leq \frac{\pi}{2}$. Figure \ref{contour} shows the numerically obtained values of $\theta_C$ in $K-J_1$ plane for various values of $J_2$. As evident from the Figure \ref{contour}, the value of $K_c$  linearly increases with $J_1$ which is expected. As one increases the values of $J_2$, $K_C$ furthers takes higher value.

Let us summarize our results for classical ground state configurations.  For all the parameter value there is a $\theta_C$ for $K \leqslant K_c$ which defines the ground state configurations according to the Eqn \ref{spincon}. The ground state has finite degeneracy in the presence of $J_0, J_1$.  For $J_1 =J_2=0$, the degeneracy if $2^{N}$ and for $J_2=0$, the degeneracy is $2^{N/2}$. For both $J_1, J_2$ nonzero, the degeneracy is reduced to 4 as described in Fig. \ref{four}.  For large $K > K_c$,  all the spins get aligned along the $x$-axis corresponding to $\theta_C= \pi/2$. Now we are in a position to discuss the relative stability of the ground state spin configurations against the quantum fluctuation as prescribed by linear spin wave theory.

\section{Linear Spin-Wave Theory}
\label{section-2}
We notice that in generals spins are quantized in an arbitrary direction which we call the local axis represented by $x^{\prime}/z^{\prime}$. The global 
axis will be represented by $x/y$. Any spins has the decomposition,

\begin{equation}
\vec{S}_{\bf{r}}=\etx S_{\bf{r}}^x+\etz S_{\bf{r}}^z 
\end{equation}

Here the index `$\bf{r}$' indicates the position of a given site.  Now we  perform an orthogonal co-ordinate transformation ( from $x,y,z$ to $x',y',z'$) such that one axis of our new co-ordinate system gets aligned along the local moment direction at every site. 
\begin{eqnarray}
\label{hptrans}
S_{\bf{r}}^x&=&S_{\bf{r}}^{x^{\prime}}\cos{\lambda_i \theta_c}-S_{\bf{r}}^{z^{\prime}}\sin{\theta_c} \\
S_{\bf{r}}^z&=&S_{\bf{r}}^{x^{\prime}}\sin{\lambda_i \theta_c}+ S_{\bf{r}}^{z^{\prime}}\cos{\theta_c} 
\end{eqnarray}
The expressions for $S^{\prime}_{x/z}$ in terms of the bosonic operators are given below,
\begin{eqnarray}
S_{\bf{r}}^{x^{\prime}}&=&s-{a^\dagger}_{\bf{r}}  {a}_{\bf{r}},~~~S_{\bf{r}}^{z^{\prime}}=\sqrt{\frac{s}{2}}\left({a^\dagger}_{\bf{r}}+{a}_{\bf{r}}\right)
\end{eqnarray}

We have specifically chosen the above representation as our interest is to investigate the phase boundary where the spins align mostly along $x$-direction.
Where $a_{\bf{r}}^{\dagger}$ $\&$  ${a}_{\bf{r}}$ represents creation $\&$ annihilation operator for magnon at  site $\bf{r}$.


\subsection{High-field limit, $K \gg J_{0},J_{1},J_{2}, \theta_0=\pi/2$}

\begin{figure}[h]
\label{highk}
\includegraphics[width=1.0\linewidth]{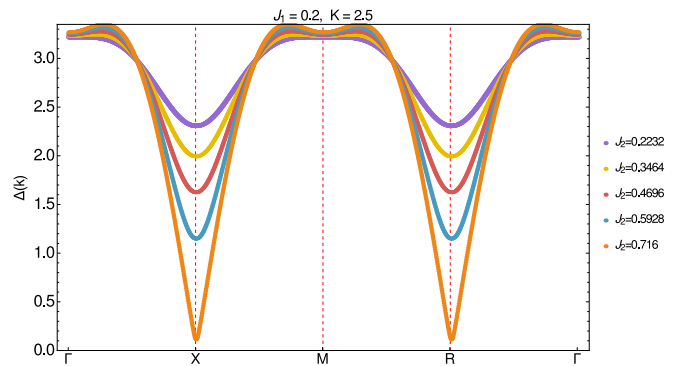}
\caption{Plot showing the dispersion in the High-field case, where quadratic
behaviour slowly vanishes to a linear behavior at the second order critical line
given by $K = 2SJ_1+\frac{29}{8}SJ_2$.} 
\end{figure}

We use Holstein-Primakoff transformation given in Eqs. \ref{hptrans} where the quantum
fluctuations are given in the form of hard-core bosons and performing the
transformation individually around the degenerate classical ground states obtained
in the previous section for different cases of $J_{2}=0$ and $J_{2}\neq 0$ in the
low-field limit we obtain the bosonic Hamiltonian given by,
\begin{eqnarray}
H &=& N\sum_{k}\big[\xi_{k}\hat{a}_{k}\hat{a}_{k}^{\dagger} + \frac{\gamma_{k}}{2}(\hat{a}_{k}\hat{a}_{-k} + \hat{a}_k^{\dagger}\hat{a}_{-k}^{\dagger})\big]  + K S N \nonumber \\
\end{eqnarray}
where,
\begin{eqnarray}
\label{gap_high}
\varepsilon_{k} &= & \gamma_{k} +  K -\frac{S J_2}{8} \nonumber \\
\gamma_{k} &= & \frac{-S}{4}\bigg[J_2(2p^2_{k}-1) - 4(J_1+J_2)p_k\bigg] \nonumber \\
\end{eqnarray}
where $p_k= \cos (k_x+ k_y) \cos (k_x- k_y)$. Diagonalizing the Hamiltonian we obtain the 
magnon-spectra as given by, 
\begin{eqnarray}
\label{dispers}
E_{k} = \sqrt{\varepsilon_{k}^{2} - \gamma_{k}^{2}}. 
\end{eqnarray}
The spectrum is plotted for few parameters as shown in Fig \ref{highk}. 
To extract the low energy behaviour of the spectrum  expand the spectrum
around the minima i.e around $X$-points as $(k_x=-\frac{\pi}{2}+\delta_x,k_y=\frac{\pi}{2}+\delta_y)$.
substituting this into Eq. \ref{dispers} and Eq. \ref{gap_high} to obtain,

\begin{eqnarray}
\label{gap_critical}
E_{\vec{\delta}}=\tilde{K}_1^{1/2} \sqrt{ \tilde{K}_2+ 4(J_1+2J_2) |\vec{\delta}|^2}
\end{eqnarray}
where $\tilde{K}_1= K - \frac{SJ_2}{8},~ \tilde{K}_2=(K-2SJ_1-\frac{29}{8}SJ_2)$.
From the above expression it is clear that the spectrum remains
gapped with quadratic low energy dispersions for all parameters 
values except for the second order critical line given by $K_c=2SJ_1+\frac{29}{8}SJ_2$
for which the spectrum is gapless and linear around minima.
 
\hspace*{-2cm}


\label{num-res}
\subsection{Low-field case, $J_{2}=0$, $K$ is finite.} 
As mentioned  before that when $K$ is small compare to other parameters of the system and $J_2=0$, the classical  ground state has huge degeneracy whose asymptotic dependence is $4^{\sqrt{\mathcal{N}_p}}$. To see their stability under quantum fluctuations, we performed the H-P transformation for each degenerate ground state and then numerically investigated whether the degeneracy is lifted and some states are selected as the ground states. We mention that these  degenerate states has random dipole moment associated with each plaquette. To get the spectrum, we use wellknown dynamical matrix method ~\cite{Colpa-1978,Ming-wen Xiao} followed by equation of motion.  The dynamical
matrix is given by,
  \begin{eqnarray}
i\frac{\partial \psi}{\partial t} & = & \big[\psi,H\big]  =  \mathcal{D}\psi 
\end{eqnarray}
\[\psi^{\dagger}  \equiv 
\begin{bmatrix}
a^{\dagger}_{1} & a^{\dagger}_{2} & \hdots & a^{\dagger}_{N} &  a_{1} & a_{2} & \hdots & a_{N} 
\end{bmatrix}  
\label{dynamicmatrix}
\] 
 \begin{eqnarray}
\mathcal{D}_{ii} & = & 2J_0 S^3\cos^4\theta + 2J_1 S \cos^2\theta + K \sin\theta \nonumber \\
\mathcal{D}_{ij} & = & -\lambda_{i}\lambda_{j} J_0\frac{S^3}{2} \sin^2\theta\cos^2\theta + \delta\frac{SJ_1}{2}\sin^2\theta
\end{eqnarray}

where, $\lambda_{i'}$s are the coefficients obtained from the mean-field analysis and $\delta=1$ if $i,j$ belong to same plaquette and are along either one of the orthogonal directions $n_1$,$n_2$ else $\delta=0$. Upon diagonalizing and picking the lowest eigenvalue we get the gap which is equal to $0.4581(4)$ for $J_0=1,J_1=0.2,K=0.2$ and $\theta=0.06691(5)$ which is in excellent agreement from the linear spin-wave calculations calculated for $J_2\neq 0$ case whose ground states are subset of the degenerate ground states (DGS) for $J_2=0$ case as explained in Fig \ref{four}.  We have found that  there is no lifting of degeneracy and the hence no order from disorder phenomenon happens here.

\hspace*{-2cm}\begin{figure}[h]
\label{four}
\includegraphics[width=0.75\linewidth]{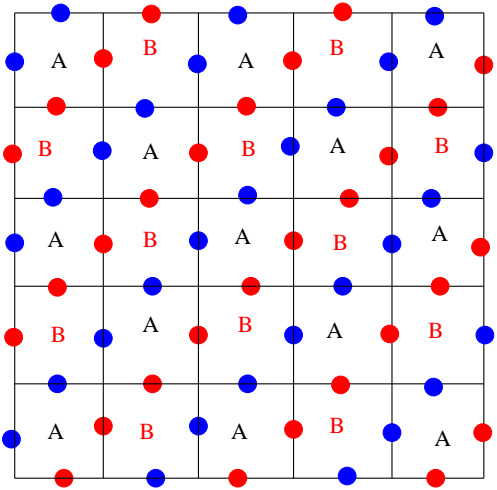}
\caption{One of the four degenerate ground states that exists for finite $J_2$ is shown in the above figure. The red dots
represents up-spins and blue dots represents down-spins. Taking the mid points of the below left square as the origin, the position
of up-spins and down spins can be written as $\vec{\mathcal{R}}_{up}=- \ety/2 + m_1 \vec{a} + n_1 \vec{b},~~\vec{\mathcal{R}}_{down}=- \etx/2 + m_2 \vec{a} + n_2 \vec{b} $ where
$\vec{a}=\etx/2 + \ety/2,~~\vec{b}=- \etx + \ety $. The 2nd degenerate ground state is obtained by replacing the up-spin by down-spins and vice versa. The third
 and fourth ground states could be obtained by rotating the spin-cpnfigurations of 1st and 2nd by $\pi/2$.} 
\end{figure}

\hspace*{-2cm}\begin{figure}[h]
\label{four}
\includegraphics[width=0.9\linewidth]{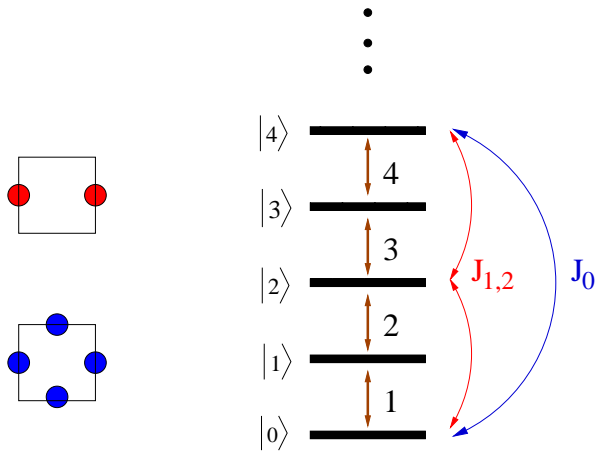}
\caption{Left column showing a cartoon picture of a  process where the action of $T_2(4)$(red(blue)) upon the ground state of 0-QP subspace producing 2-QP(above) and 4-QP(below) is shown. The right column shows a simple connection of the perturbing parameters connecting the different QP levels of the unperturbed Hamiltonian.} 
\end{figure}

\label{num-res}
\subsection{Low-field case, $J_{2}\neq 0$}

Similarly to previous sections here we perform a HP transformation around the ground states of ordered dimer coverings shown in Fig \ref{four}. 
\begin{eqnarray}
H_{i} &=& N\sum_{k}\bigg[\varepsilon_k a^{\dagger}_k a_k + \frac{\tilde{\gamma}_{i,k}}{2}( a^{\dagger}_{k} a^{\dagger}_{-k} + a_{k} a_{-k})\bigg] \nonumber \\
      &&  - \frac{1}{2} J_0 S^4 N \mathcal{C}_{\theta}^4 - (J_1  + 2 J_2 ) S^2N  \mathcal{C}_{\theta}^2
\end{eqnarray} 
where $\varepsilon_k$ and $\gamma_{i,k}$ is given below,
\hspace*{-2cm}\begin{figure}[h]
\label{lowfield}
\includegraphics[width=1.05\linewidth]{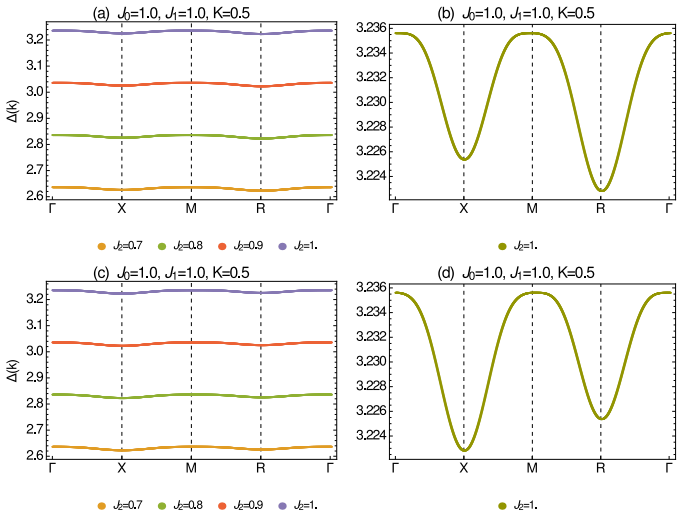}
\caption{In the above figure we have plotted the spin-wave dispersion for  low values of $K$. The upper right panel (b) shows the spectrum for 1st and 2nd degenerate ground states as mentioned in Fig. \ref{four}. Similarly lower right panel (c) shows the spectrum for 3rd and 4th ground states. The upper (a) and lower left columns (d) shows the variations of spectrum for different values of $J_2$. } 
\end{figure}

\begin{eqnarray}
\varepsilon_{k} &= & \gamma_{k} + 2S^3J_{0} \mathcal{C}_{\theta}^4  + 2S J_{1}\mathcal{C}_{\theta}^2  -\frac{J_2S}{8} \mathcal{S}_{\theta}^2  \\
\tilde{\gamma}_{i,k} &= & \mathcal{S}_{\theta}^2\bigg[\gamma_k + S^3J_0\mathcal{C}_{\theta}^2 [2 \chi_i q_k + p_k] \bigg]
\end{eqnarray}
In the above  and  $i=1,2,3,4$ corresponds to four degenerate ground states explained in Fig \ref{four}. The value of $\chi_i=1(-1)$ for $i=1,2(3,4)$ and $p_k, q_k, s_k$ are
defined below. In the above $\mathcal{C}_{\theta}$ and $\mathcal{S}_{\theta}$ stands for $\cos \theta$ and $\sin \theta$ respectively and $q_k= \sin k_x  \sin k_y $ 
and $p_k$ is defined after Eq [17].The spectrum is found gapped for all values of $K, J_2$ and given by 

\begin{eqnarray}
\Delta_{k} &= & \zeta_0 \sqrt{\zeta_1 + \zeta_2 \varrho^2} \nonumber
\end{eqnarray}
where $\zeta_0=(2S^3J_{0}\cos^{4}\theta_{0} + 2S( J_{1} + 2J_2) \cos^{2}\theta_{0} -\frac{SJ_2}{8} \sin^2 \theta_0)^{0.5}$, $\zeta_1=2J_1(S\cos^2\theta_0 - \frac{S}{2}\sin^2\theta_0) + 8S^3J_0\cos^4\theta_0 - (2\chi +1)\frac{S^3J_0}{2}\cos^2\theta_0\sin^2\theta_0-\frac{15S}{8}J_2\sin^2\theta_0$ and $\zeta_2=\frac{\sin^2\theta_0}{2}\bigg[ 4S(J_1+3J_2) + 2S^3\chi J_0\cos^2\theta_0 \bigg]$. The constant of relevance here $\zeta_1$ was solved numerically and simultaneously with the equation that minimizes the total energy given in eqn.\eqref{clasen}. The results for few parameter values are shown in Fig. \ref{low-field} and we see the quadratic behavior for all the parameters values about the X,R-High symmetry points for ground states oriented along the $\pm x$, $\pm y$ respectively. 


\section{PCUT}
\label{section-3}
The Hamiltonian as represented by Eq. \ref{ham} and  equations following that represent in general a quantum interacting spin 
system which is often not solvable exactly. Generally one follows various analytical schemes depending on their 
interest of specific aspect of the model or the suitability of the model to the method itself. For example for one dimensional
nearest neighbour interacting spin system Jordan-Wigner transformation~\cite{fradkin-1989} is a very good starting point for
analytical solutions. For 2 dimensional and in higher dimensional system in general slave fermion/boson methods are
applied and often meanfield approximations are followed ~\cite{wen-2002}. Though there are exceptions for example 
Kitaev model ~\cite{kitaev-2006,kitaev-2003} which represents an exactly solvable model. As mentioned before, in this article we use PCUT 
(perturbative continuous unitary transformation) in our model whenever applicable. We are motivated by the fact that 
unlike slave boson/fermion formalism, we do not have to use meanfield approximations which is often does not represents 
the true ground states, on the otherhand few properties like gap and one particle spectrum can be calculated to very high 
order of perturbations  as shown in previous studies~\cite{vidal-2009}. Another important aspects of PCUT is the sequential
derivations of effective Hamiltonian at higher order  within a given particle sectors. The pre-requisite of applicability of  PCUT is the 
equidistant spectrum which we have in the limit   $K\gg J_0,J_1,J_2$ where the ground state is given by the spins aligned 
align the $x$-axis and excitations are given by spin flip excitations. We use PCUT to obtain the accurate measure of  
zero-particle gap and one particle dispersion  and effective Hamiltonian  upto reasonably high order. We  compare our results of 
gap with previous monte-carlo  studies~\cite{Chyh-Hong Chern} and find very good comparison. Our one particle dispersion should 
also be useful to verify the low energy excitations beyond spin-wave approximations. Before we start the application of PCUT we first map the spin-Hamiltonian \eqref{ham} onto the the effective bosonic operator formalism. Since the 
ground state in the high-field limit is where all the spins are along $x$-axis,  we transform our global $z$-basis along the ground 
state orientation of spin as: 

\begin{eqnarray}
\label{rot_spin}
{S^x}_{\bf{r}}&=&{S^{x'}}_{\bf{r}}\cos{ \theta}-{S^{z'}}_{\bf{r}}\sin{\theta} \\
{S^z}_{\bf{r}}&=&{S^{x'}}_{\bf{r}}\sin{ \theta}+{S^{z'}}_{\bf{r}}\cos{\theta} 
\end{eqnarray}
where $\theta=\pi/2$, the expressions for $S^{\prime}_{x/z}$ in terms of the bosonic operators are given below,
\begin{eqnarray}
\label{boson_spin}
{S^{x}}_{\bf{r}}&=&s-{a^\dagger}_{\bf{r}}  {a}_{\bf{r}},~~~{S^{z}}_{\bf{r}}=\sqrt{\frac{s}{2}}\left({a^\dagger}_{\bf{r}}+{a}_{\bf{r}}\right)
\end{eqnarray}
We now proceed to establish PCUT for our Hamiltonian \eqref{ham} using Eqs \ref{rot_spin}. The resulting Hamiltonian has quadratic 
and as well quartic interaction which amounts to an interacting Hamiltonian. Our aim   is to transform the resulting Hamiltonian  to an unitarily 
equivalent Hamiltonian ($H_{\text{eff}}$) through a unitary transformation. Specifically, We follow the reference \cite{F. Wegner} where the 
block-band diagonality of the Hamiltonian is preserved with the choice of quasi-particle (QP) conserving infinitesimal generator $\eta(\ell)$ 
given by Wegner. We request the reader to follow the references for further pedagogical review. In the bosonic operator representation, 
the initial Hamiltonian \eqref{ham} takes the form:
\begin{eqnarray}
\label{hpcut}
H & = & -\frac{N}{2} + Q + T_{-4} + T_{-2} + T_{0} + T_{2} + T_{4}
\end{eqnarray}
where, $T_{4}$ $\propto$ $J_0$, $T_{2}$ $\propto$ $J_1,J_{2}$ are the effective operators corresponding to the perturbed Hamiltonian in the bosonic operator representation and $Q$ is unperturbed Hamiltonian whose eigenstates serve as the basis for the Hilbert space of our system with $\big[Q,T_n\big]=nT_n$. Using eqns. \eqref{rot_spin}, \eqref{boson_spin} the exact expressions for the  effective operators $T_n$ can be found as,

\begin{eqnarray}
\label{hopping}
T_0 & = & J_0\sum_{\square}\big[a_1a_2a_3^{\dagger}a_4^{\dagger} + a_1a_2^{\dagger}a_3a_4^{\dagger} + a_1a_2^{\dagger}a_3^{\dagger}a_4 + \text{h.c}\big] \nonumber \\
&& + J_1\sum_{\square}\big[a_1a_3^{\dagger} + a_2a_4^{\dagger} + \text{h.c}\big] \nonumber \\
&& +\frac{J_2}{8}\sum_{\langle A B \rangle}\big[a_2a_4^{\dagger} + a_3a_7^{\dagger} + a_1a_5^{\dagger} + a_2a_6^{\dagger} - a_3a_7^{\dagger} \nonumber \\ 
&& - a_1a_7^{\dagger} - a_4a_6^{\dagger} + \text{h.c}\big] \nonumber \\
T_2 & = & J_0\sum_{\square}\big[a_1^{\dagger}a_2a_3^{\dagger}a_4^{\dagger} + a_1^{\dagger}a_2^{\dagger}a_3a_4^{\dagger} + a_1^{\dagger}a_2^{\dagger}a_3^{\dagger}a_4 + a_1a_2^{\dagger}a_3^{\dagger}a_4^{\dagger}  \big] \nonumber \\
&& + J_1\sum_{\square}\big[a_1^{\dagger}a_3^{\dagger} + a_2^{\dagger}a_4^{\dagger} \big] \nonumber \\
&& +\frac{J_2}{8}\sum_{\langle A B \rangle}\big[a_2^{\dagger}a_4^{\dagger} + a_3^{\dagger}a_7^{\dagger} + a_1^{\dagger}a_5^{\dagger} + a_2^{\dagger}a_6^{\dagger} - a_3^{\dagger}a_7^{\dagger} \nonumber \\ 
&& - a_1^{\dagger}a_7^{\dagger} - a_4^{\dagger}a_6^{\dagger} \big] \nonumber \\
T_4 & = & J_0\sum_{\square}\big[a_1^{\dagger}a_2^{\dagger}a_3^{\dagger}a_4^{\dagger} \big] \nonumber \\
\end{eqnarray}

where the numbering is according to FIG.\ref{squareacid}c. The conjugate operators can be found as $T_{n}^{\dagger}=T_{-n}$. With the effective operators in hand, the Hamiltonian \eqref{hpcut} can now be transformed  to an effective Hamiltonian $H_{\text{eff}}=U^{\dagger}HU$ commuting with $Q$. The infinitesimal unitary transformation is given by, 
$U \approx 1+i \eta(\ell)$ where  $\partial_\ell H=\big[\eta(\ell),H(\ell)\big]$ which determine the effective Hamiltonian in the limit 
$\ell \rightarrow \infty$. The above choice of generator is given in the pioneering work of Wegner~\cite{F. Wegnero,F. Wegner}. Starting from the 
Hamiltonian \eqref{hpcut}, we get the block-band diagonal effective Hamiltonian perturbatively  order by order in our case. The general form of effective Hamiltonian can be represented as~\cite{C. Knetter,C. Knettero}:
\begin{eqnarray}
\label{heff}
\begin{split}
H_{\text{eff}} = -\frac{N}{2} + Q + \sum_{k=1}^{\infty}\sum_{\substack{|m|=k, \\ M(m)=0}}C(\underline{m})T_{m_{1}}\cdot\cdot\cdot T_{m_{k}}
\end{split}
\end{eqnarray}
where the series expansion coefficients $c(\underline{m})$ are computed before hand using the flow equations. The next job in solving the Hamiltonian \eqref{heff} is to determine it's action in each subspace of a given QP number $q$ and diagonalize $H_{\text{eff}}$ in each of these subspaces. In the following sections we computed the above Hamiltonian in the respective subspaces of $q=0,1$ and we present the results of ground state energy per spin $(q=0)$ and one particle gap $(q=1)$ in the confined phase($J_2=0$) and paraelectric phase($J_2\neq 0$) respectively.

\label{pcut-j2=0}
\subsection{Large-field limit ($J_{2} = 0, K \gg J_{1},J_{0}$)}

As mentioned in the outline, to determine the value of $K_{c}$ we perform the PCUT method considering the low-energy sector of the Hamiltonian $(1)$ in the strong-field limit $K\gg J_{0},J_{1},J_{2}=0$  of CDT. The ground state(0-QP state) is fully polarized in the x-direction and the elementary excitations(1-QP state) are the single spin flips with energy cost $2K$, setting K=1/2 at order 7 we obtain the ground state energy and one-particle dispersion given by,
\begin{align}
\begin{split}
e_{0} & = -\frac{1}{2} -\frac{1}{8}J_{0}^2 - \frac{1}{384}J_{0}^4 - \frac{41}{393216} J_{0}^6 - \frac{1}{2}J_{1}^2 \\
&\quad- \frac{1}{2}J_{0}J_{1}^2 - \frac{9}{32} J_{0}^2 J_{1}^2 - \frac{3}{16} J_{0}^3 J_{1}^2 - \frac{47525}{442368} J_{0}^4 J_{1}^2  \\
&\quad- \frac{1}{8}J_{1}^4 - \frac{1}{2}
 J_{0} J_{1}^4 - \frac{32957}{32768}J_{0}^2 J_{1}^4 - \frac{61}{512} J_{1}^6
\end{split}
\end{align}
\begin{align}
\begin{split}
\Delta & = 1 - \frac{1}{2}J_{0}^2 + \frac{3}{32}3 J_{0}^4 - \frac{1711}{27468}J_{0}^6 + 2 J_{0} J_{1}  + 2 J_{1} \\
&\quad - \frac{1}{8}
 J_{0}^2 J_{1} - \frac{3}{4} J_{0}^3 J_{1} + \frac{67}{576}J_{0}^4 J_{1} + \frac{8131}{13824}J_{0}^5 J_{1} \\
 &\quad - \frac{3}{4}
 J_{0}^2 J_{1}^2 - \frac{7}{2} J_{0}^3 J_{1}^2 - \frac{111013}{36864}J_{0}^4 J_{1}^2 + J_{0} J_{1}^3  \\
&\quad + \frac{317}{128}
 J_{0}^2 J_{1}^3 - \frac{13361}{4608} J_{0}^3 J_{1}^3 - \frac{10207}{3072}J_{0}^2 J_{1}^4 + \frac{79}{128} J_{0} J_{1}^5
\end{split}
\end{align}

\begin{figure}[h]
\label{pcut-spectrum1}
\includegraphics[width=1.0\linewidth]{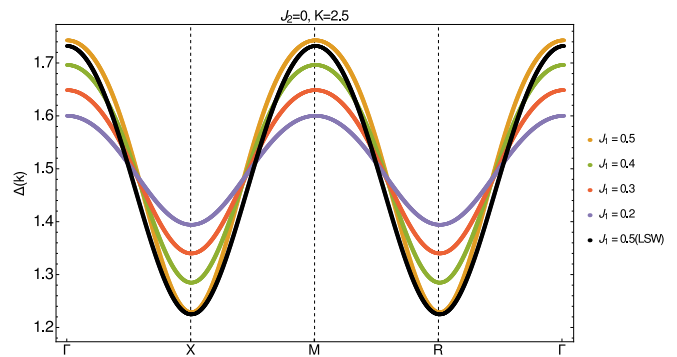}
\caption{Here we have plotted the spectrum as obtained from PCUT upto 7th order. The black line is the result obtained from spin-wave
approximation.} 
\end{figure}

\begin{figure}[h]
\label{phase-diag}
\includegraphics[width=1.0\linewidth]{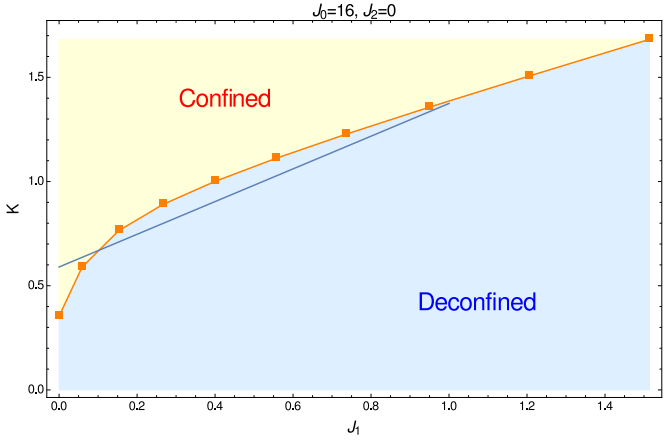}
\caption{The above figure shows the phase boundary between confined and deconfined phase as obtained by PCUT. The solid blue line
is a linear regression fit of our PCUT result. } 
\end{figure}
A standard analysis for the gap $\Delta$ using DLogPade approximants suggests that below $J_{0}=0.8$ there were no closure of the gap. Suggesting the quantum phase transition below $J_{0}=0.8$ may not be of second order. However at $J_{0}=1.5$ we obtain the relation of critical line $K_{c}=2.931(5)J_{1}$.  In Fig. \ref{phase-diag}, we presented the phase diagram from confined to deconfined phas transition in $K-J_1$ plane. In Fig. \ref{pcut-spectrum1}, we have plotted the dispersion as obtained from PCUT and compared with spin wave spectrum obtained earlier which matches each other very well.

\label{num-res}
\subsection{Large-field limit ($J_{2} \neq 0, K \gg J_{2},J_{1},J_{0}$)}
In this section, we present the results of ground state energy per spin and the one particle dispersion for $J_2 \neq 0$. We could perform it till 4th order as the number of intermediate states increase due to non-local nature of dipole-dipole interaction. Nevertheless, the obtained expression are helpful in determining it's critical behavior qualitatively. As in the above case the ground state is the state where all the spins are oriented along field-direction and the single-spin flips are the elementary excitations with $2K$ cost of energy, hence setting $K=1/2$ we obtain the ground state energy and one-particle dispersion as,     

\begin{eqnarray}
\label{e0-j2neq0}
e_{0}&=&-\frac{1}{8}J_{0}^2 - \frac{1}{384}J_{0}^4 - \frac{1}{2}J_{1}^2 - \frac{1}{2}J_{0} J_{1}^2 - \frac{9}{32} J_{0}^2 J_{1}^2 - \frac{1}{8}J_{1}^4 \nonumber \\ 
&& - \frac{1}{2}J_{1} J_{2} - \frac{1}{2}J_{0} J_{1} J_{2} - \frac{9}{32}J_{0}^2 J_{1} J_{2} - \frac{3}{8} J_{1}^2 J_{2} - \frac{3}{4} J_{0} J_{1}^2 J_{2} \nonumber \\
&& - \frac{1}{4}
 J_{1}^3 J_{2} - \frac{3}{8}3 J_{2}^2 - \frac{1}{8}J_{0} J_{2}^2 - \frac{39}{256} J_{0}^2 J_{2}^2 - \frac{33}{32}  
 J_{1} J_{2}^2 \nonumber \\
&& - \frac{3}{2} J_{0} J_{1} J_{2}^2 - \frac{155}{64} J_{1}^2 J_{2}^2 - \frac{15}{32} J_{2}^3 - \frac{9}{16}
 J_{0} J_{2}^3 \nonumber \\
&&  - \frac{207}{64}J_{1} J_{2}^3 - \frac{2859}{2048} J_{2}^4  \\
\label{delta0-j2neq0}
\Delta & = &1 +2J_{1}
-\frac{1}{2}J_{0}^2 + \frac{3}{32} J_{0}^4 + 2 J_{0} J_{1} - \frac{1}{8}J_{0}^2 J_{1} - \frac{3}{4} J_{0}^3 J_{1}\nonumber \\
&& - \frac{3}{4}
J_{0}^2 J_{1}^2 + J_{0} J_{1}^3 + J_{0} J_{2} - \frac{3}{8} J_{0}^3 J_{2} + J_{1} J_{2} + \frac{5}{2} J_{0} J_{1} J_{2} \nonumber \\
&&+ 
 \frac{105}{64} J_{0}^2 J_{1} J_{2} + \frac{3}{4}J_{1}^2 J_{2} + 3 J_{0} J_{1}^2 J_{2} + \frac{1}{2}J_{1}^3 J_{2} \nonumber \\
&& + \frac{11}{8} J_{2}^2 + \frac{41}{16}
 J_{0} J_{2}^2 - \frac{29}{128}J_{0}^2 J_{2}^2 + \frac{45}{16} J_{1} J_{2}^2 \nonumber \\
&& + \frac{233}{16} J_{0} J_{1} J_{2}^2 + \frac{71}{8} J_{1}^2 J_{2}^2 + \frac{9}{4}J_{2}^3 + \frac{297}{32} J_{0} J_{2}^3 \nonumber \\
&& + \frac{535}{32}J_{1} J_{2}^3 + \frac{553}{64}
 J_{2}^4
\end{eqnarray}

\begin{figure}[h]
\label{pcut-spectrum2}
\includegraphics[width=1.0\linewidth]{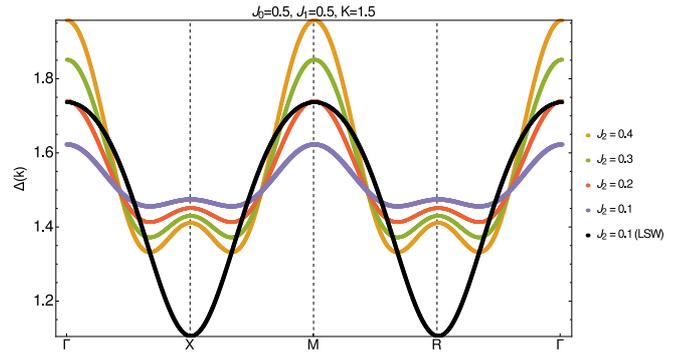}
\caption{In the above we have plotted one particle dispersion as obtained from PCUT. The above results is also compared with
the spin-wave spectrum (the black plot). The additional peaks in the PCUT results suggests appearance of para-electric phase~\cite{Chyh-Hong Chern}
for the presence of finite $J_2$ } 
\end{figure}

Due to the increasing demand of computing power, we could only perform PCUT upto fourth order and find that gap does not  vanishes
unlike $J_2=0$ case. However this is fully consistent to that fact that switching on $J_2$  triggers a para-electric phase as obtained
in presvius study \cite{Chyh-Hong Chern} which has excitations as a single dipole changing its direction costing a finite gap. Also
in Fig. \ref{pcut-spectrum2}, we have plotted the dispersion as obtained from PCUT. We see that additional peaks appears
in the dispersions when compared to spin-wave dispersions. This additional peaks denotes the emergence of paraelectric phase.
\begin{figure}[h]
\label{pcut-3D}
\includegraphics[width=0.5\linewidth]{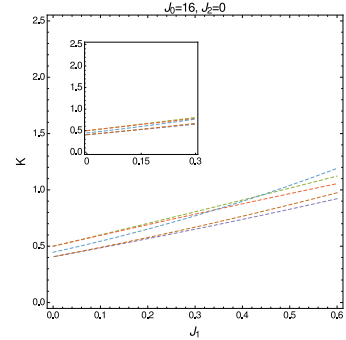}
\caption{ The CDT transition in $K-J_1$ plane as obtained in various order of PCUT. The straight line which lie above
are obtained in low order of PCUT. We see that at higher order it decreases the deconfined phase suggesting the success of PCUT
to estimate the one particle gap closure effectively.} 
\end{figure}

\begin{figure}[h]
\label{pcut-3D}
\includegraphics[width=1.0\linewidth]{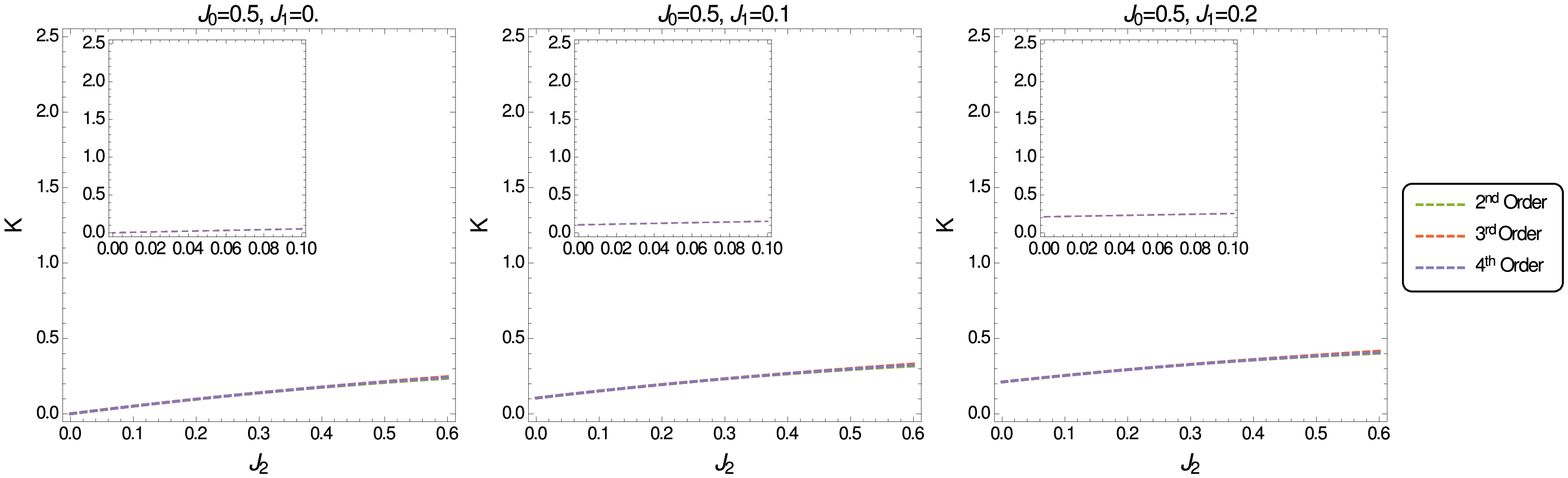}
\caption{ The CDT as obtained for finite $J_2$. We mention that the gap does not vanish really and the above plot is obtained
from eq. \ref{delta0-j2neq0} by solving for $K_c$ from $\Delta=0$. } 
\end{figure}

Our results for gap and 1 particle dispersion will have experimental consequences and can be verified. Below we discuss the model in the limit $J_0 >> K$.
\section{Dual mapping and the extended model}
\label{section-4}
As computed in a previous study~\cite{Bo-Jie Huang and Chyh-Hong Chern} the model is shown to map to the celebrated Toric Code Model (TCM) in the low-field limit $J_0\gg K_x$ and $J_{1,2}=0$. Here for brevity we show the duality of the extended model where an extra field $K_y\ll J_0$ is applied along the transverse direction and show how it can be mapped to an interacting anyon model discussed in previous work~\cite{vidal-2009}. 

\begin{eqnarray}
\label{duality_H}
H & = & -J_0\sum_{\square}\sigma_1^z\sigma_2^z\sigma_3^z\sigma_4^z -K_x\sum_{i}\sigma_1^x  -K_y\sum_{i}\sigma_1^y \nonumber \\
\end{eqnarray}

In the limit $J_0\gg K_x,y$ the Hamiltonian \eqref{duality_H} can be transformed to an a transverse field model with a simple z-rotation . The new Hamiltonian in the transformed variables is given as,

\begin{eqnarray}
\label{red_duality_H}
H & = & -J_0 \sum_{\square}\sigma_1^{z^{\prime}} \sigma_2^{z^{\prime}} \sigma_3^{z^{\prime}} \sigma_4^{z^{\prime}} -K_x^{\prime}\sum_{i}\sigma_1^{x^{\prime}} \nonumber \\
\end{eqnarray}

where $K_x^{\prime}=(1-\tan\theta)K_x$. Now for $J_0 > K_x^\prime$  the Hamiltonian can be split as, $H=H_0+H_1$ limit with the standard perturbation theory of using the projector operator formalism the effective Hamiltonian at $4^{\text{th}}$ is
\begin{eqnarray}
\label{pert}
H_{\text{eff}}^{(4)} & = & \mathcal{P}(H_{1}\mathcal{D})^3 H_1\mathcal{P} \nonumber \\
\mathcal{D} && = -\frac{1-\mathcal{P}}{H_0-E_0}
\end{eqnarray}
using the eqns. \eqref{red_duality_H},\eqref{pert} the effective Hamiltonian can be found in our case as, 

\begin{eqnarray}
\label{new_H}
H_{\text{eff}}^{(4)} & = & -J_0 \sum_{\square}\sigma_1^{z} \sigma_2^{z} \sigma_3^{z} \sigma_4^{z} -\tilde{K}\sum_{+}\sigma_\alpha^{x} \sigma_\beta^{x} \sigma_\gamma^{x} \sigma_\delta^{x}\nonumber \\
\end{eqnarray}

where $\tilde{K}=\frac{5K_x^{\prime}}{16J_0^3}$ and the summation  $+$ runs over all the vertices. The Hamiltonian \eqref{new_H} is nothing but the Toric Code Hamiltonian \cite{kitaev-2003}. In order to be in the deconfined($K<0.325J_0$) phase the coupling strengths should satisfy $\frac{5{K_x^{\prime}}^5}{16J_0^4}<0.325$.  This mapping establishes a  connection starting from Ising lattice gauge theory to Kitaev Hamiltonian ~\cite{kitaev-2003}. At fourth order, the anyons are non-interacting but beyonds 4-th order they becomes interacting as shown before ~\cite{vidal-2009}. The implication of $J_1$ and $J_2$ interaction on such interacting anyonic system is left as an future study.

\section{Discussions} 
\label{discussion}

To summarize, in this study we have performed first ever analytical study of a model $\rm{H_2SQ}$ system. Our work builds on the model proposed earlier~\cite{H.-D. Maier,Chyh-Hong Chern}. To start with, we have clearly mentioned the various terms in the Hamiltonian and its physical origin. At the zeroth level, the model Hamiltonian has only plaquette term which harbors a deconfined phase. The application of an external magnetic field (given by a strength $K$ ) drives the deconfined phase to a confined phase. We have determined the value of $K_c$ for which such transition happens. The role of intermolecular-coupling $J_{1}$ and dipole-dipole interaction $J_{2}$ on such transition has been investigated. We have shown that the role of $J_1$ and $J_2$ is to stabilize the deconfined phase. The ground state without dipole-dipole interaction term in the low field case was found to be  singlet pair or dimers whose z-component projection of spin are anti-aligned to satisfy "ice-rules". Thus the classical ground state remains degenerate for small values of $K$.In the presence of the dipole-dipole term $J_{2}$ the local degeneracy has been  removed to yield a  four degenerate global ground states independent of system size.  The role of quantum fluctuation
has been investigated over these large classical degenerate ground states. Surprisingly we have found that, at quadratic level, the local degeneracy is not removed and there is no order from disorder phenomena occurs. We have found the spin-wave dispersion for the four global degenerate ground states and found
that though in general the spectrum is gapped and quadratic for small values of $k$, for certain parameters, the spectrum becomes gapless and liner. This happens particularly near the phase boundary of confinement to deconfinement phase transition. Our formula for spin-wave dispersion and measure of gap will
be useful for future experiment. To give a more meaningfulness to our study, we have applied PCUT to analysis the system in the large field limit ($K >> J_0, J_1, J_2$) where the ground state consists of all spin aligned along $x$-direction and excitation consists of spin flip excitations. We have improved
the estimation of ground state energy and one particle gap as estimated from spin-wave analysis.  However we believe that  further improvement of our work on PCUT can be done and it is left as a future study.

\section{Acknowledgments}
We thank prof Julien Vidal for sharing his expertise on PCUT with us. We also thank Prof A. M.  Jayannavar for encouraging us in the initial stages of this work.

\end{document}